\documentclass[prb,preprint,showpacs,eqsecnum]{revtex4}
\usepackage{graphicx}
\begin{document}
\title{Ground state, vibrational spectrum and deterministic
transport of a chain of charged particles}
\author{S.~I.~Denisov}
\email{denisov@ssu.sumy.ua}
\author{E.~S.~Denisova}
\affiliation{Department of Mechanics and Mathematics,
Sumy State University, 2, Rimskiy-Korsakov Street, 40007 Sumy,
Ukraine}

\begin{abstract}
We study the equilibrium, oscillatory and transport properties of
a chain of charged particles which interact with each other via
the Coulomb and power-like repulsive interactions. Exact,
analytical expressions for the energy of the ground state,
interparticle distances, and vibration spectrum are derived, and
the stability criterion for a chain with equidistant particles is
established. We also consider the phenomenon of deterministic
transport of a chain in a ratchet potential subject to a
longitudinal alternating electric field.
\end{abstract}


\maketitle

\section{Introduction}

A chain of coupled oscillators is a universal mathematical model
that describes a large variety of phenomena in physics, chemistry,
biology and other sciences. Examples include energy localization
and propagation,\cite{Ford92,SzLa98,ReSaLi01} discrete
solitons,\cite{FlWi98,HeTs99} chemical reactions,\cite{Ku84}
nonequilibri\-um phase transitions,\cite{BrPaTo94,LiHu96}
stochastic resonance,\cite{Li...95,HaKiKo98}
syn\-chronization,\cite{PiRoKu01} brain dynamics\cite{Haken02} and
many others. A special case of that model, a chain of charged
particles that interact via the Coulomb interaction, also has a
variety of applications. In particular, if the charges have the
same sign and are confined by a trapping potential, it describes
the so-called one-dimensional Coulomb crystals, i.e., a
crystallized structure of ions in a linear trap formed by external
focusing fields.\cite{Du97,DuNe99,MoEs01} If neighboring charges
have opposite signs and a repulsive interaction exists between
particles, then a such chain represents a relevant model for
one-dimensional ionic crystals. It seems that those models can be
useful also for the study of one-dimensional crystals within
carbon nanotubes.\cite{Hi...00,Me...00}

During the last decade, much attention was devoted to the transport
properties of particles in thermal ratchets, i.e., spatially
periodic stochastic models exhibiting noise-induced macroscopic
transport.\cite{Ma93,AsBi94,DoHoRi94,Do95,BaReHa96,As97,Pa98,%
Po...00,CaWu01,Re02} The ratchet effect or stochastic transport in
asymmetric potentials is important for biological
motors,\cite{Ho98,VaMi00} separation techniques,\cite{Ro...94}
surface smoothing\cite{DeLeBa98} and vortex density
reduction,\cite{Le...99} to name a few. Most studies of ratchets,
including deterministic ones,\cite{SaLa99,Ma00} i.e., ratchets
without internal and external noises, have been performed for a
\textit{single} particle. However, examples of coupled
particles\cite{CsFaVi97,Bu...00} and vortices\cite{Ol...01} show
that their \textit{collective} interactions can significantly
modify the transport properties from the single particle case. To
our knowledge, the transport properties of a chain of strongly
coupled charged particles in deterministic ratchets have not been
studied at all.

In this paper, we study the equilibrium, oscillatory and transport
properties of a chain where the neighboring particles have the
same mass and opposite charges. Particles interact with each other
via both the Coulomb interaction and a repulsive interaction that
depends on the interparticle distance as $|x_{i}-x_{j}|^{-r}$
($x_{i}=x_{i}(t)$ is the coordinate of the $i$th particle, $r>1$).
In Sec.~II, we consider the ground state of such a chain depending
on the exponent $r$. Three different states of a chain, namely, a
periodic chain with equidistant particles, a periodic chain with
non-equidistant particles, and a chain with infinite period, are
analyzed, and the interparticle distances and the chain energies
as functions of $r$ are calculated. In contrast to intuitive
expectations, we show that there is a critical value of the
exponent, $r=r_{2}\approx2.799$, such that for $r>r_{2}$ the
minimum energy occurs for a chain with equidistant particles, and
for $r<r_{2}$ the minimum energy occurs for a chain with infinite
period.

Propagation of longitudinal waves in a chain with equidistant
particles, its stability with respect to small displacement of
particles from their equilibrium positions, and forced vibrations
of particles under action of an alternating electric field are
studied in Sec.~III. We derive the set of equations that describe
small oscillations of the particles, find the phonon spectrum and
its asymptotics near the center and boundary of the Brillouin
zone, and calculate the amplitude of forced vibrations. We show
that the spectrum consists of one branch; on the boundary of the
Brillouin zone it has a non-analyticity which is caused by the
long-range action of the Coulomb interaction, and a chain is
stable only if $r>r_{1}\approx 2.779$.

The deterministic transport of a chain is studied in Sec.~IV. We
consider the case when the ratchet potential is charge
independent, and a chain is driven by a longitudinal alternating
electric field. We find the average velocity of a chain and
analyze in detail its dependence on the parameters characterizing
the interparticle interactions, ratchet potential and alternating
electric field. Concluding remarks are contained in Sec.~V, and
some technical details are given in the Appendixes.

\section{Ground state}

We consider a chain of particles which have the same mass $M$ and
any neighboring particles have opposite charges, i.e.,
$q_{i+1}=-q_{i}$ and $|q_{i}|=q$ (the index $i$ labels the
particles). We assume that in addition to the Coulomb interaction
a repulsive interaction exists between the particles. In analogy to
ionic crystals,\cite{AsMe76} we choose a repulsive interaction
that depends on a power of the interparticle distance
$|x_{i}-x_{j}|$, and write the total interaction energy of a chain
in the form
\begin{equation}
     W_{int}=\frac{1}{2}{\sum_{i,j}}'\frac{q_{i}q_{j}}{|x_{i}-
     x_{j}|}+\frac{b}{2}{\sum_{i,j}}'\frac{1}{|x_{i}-x_{j}|^{r}}.
     \label{eq:energy_gen1}
\end{equation}
Here $b$ is a dimensioned parameter characterizing the strength of
the repulsive interaction, and the primes on the summation signs
indicate the absence of self-interaction terms, i.e., $i\neq j$.
To prevent the unbounded contraction of a chain, we require
$r>1$.

If a chain is periodic in equilibrium (see Fig.~1), then the
coordinates of the positive and negative charges are written as
$np+x_{n}^{+}$ and $np+a+x_{n}^{-}$, respectively. Here $p$ is the
chain period, $n$ $(n=0,\pm1,..., \pm\infty)$ numbers the chain
cells which contain two particles, $a$ is the interparticle
equilibrium distance in the cells $(a<p)$, and
$x_{n}^{+}=x_{n}^{+}(t)$ and $x_{n}^{-}=x_{n}^{-}(t)$ are the
displacements of positive and negative charges from their
equilibrium positions. With these definitions
Eq.~(\ref{eq:energy_gen1}) reads
\begin{eqnarray}
     W_{int}&=&-\frac{q^{2}}{2}{\sum_{n,m}}'\left[
     \frac{2}{|(n-m)p-a+x_{n}^{+}-x_{m}^{-}|}\right.
     \nonumber\\
     &&-\frac{1}{|(n-m)p+x_{n}^{+}-x_{m}^{+}|}
     \nonumber\\
     &&\left.-\frac{1}{|(n-m)p+x_{n}^{-}-x_{m}^{-}|}\right]
     \nonumber\\
     &&-q^{2}\sum_{n}\frac{1}{|a-x_{n}^{+}+x_{n}^{-}|}
     \nonumber\\
     &&+\frac{b}{2}{\sum_{n,m}}'\left[\frac{1}{|(n-m)p+x_{n}^{+}
     -x_{m}^{+}|^{r}}\right.
     \nonumber\\
     &&\left.+\frac{1}{|(n-m)p+x_{n}^{-}-x_{m}^{-}|^{r}}\right]
     \nonumber\\
     &&+b\sum_{n,m}\frac{1}{|(n-m)p-a+x_{n}^{+}-x_{m}^{-}|^{r}},
     \label{eq:energy_gen2}
\end{eqnarray}
where summation on $n$ and $m$ is taken over all integers. Note
also that the particles cannot move past each other, and therefore the
conditions $a>x_{n}^{+}-x_{n}^{-}$ and $p-a>x_{n}^{-}-x_{n+1}^{+}$
must hold for all $n$ and $t$.

In equilibrium $x_{n}^{+}=x_{n}^{-}=0$, and
Eq.~(\ref{eq:energy_gen2}) is reduced to the form
\begin{eqnarray}
     W_{int}^{(0)}&=&-N\frac{q^{2}}{2a}-N\frac{q^{2}}{2p}\sum_{n=1}^
     {\infty}\left[\frac{1}{n-\alpha}+\frac{1}{n+\alpha}-
     \frac{2}{n}\right]
     \nonumber\\
     &&+N\frac{b}{p^{\,r}}\zeta(r)+N\frac{b}{2p^{\,r}}
     \sum_{n}\frac{1}{|n+\alpha|^{r}},
     \label{eq:energy_equil1}
\end{eqnarray}
where $N$ is the number of particles in a chain $(N\to\infty)$,
$\alpha=a/p$ $(0\leq\alpha<1)$, and $\zeta(r)=\sum_{n=1}^{\infty}
n^{-r}$ is the Riemann zeta function. Taking into account that the
energy $W_{int}^{(0)}$ is invariant with respect to the
replacement of $\alpha$ by $1-\alpha$, we conclude that such
chains are equivalent. Therefore, the permissible values of
$\alpha$ can be restricted by the condition $0\leq\alpha\leq1/2$.

Defining the dimensionless parameter $\beta=p/l$ $(0<\beta \leq
\infty)$ and the length scale $l=(b/q^{2})^{1/(r-1)}$, we can
rewrite Eq.~(\ref{eq:energy_equil1}) as $W_{int}^{(0)}=
N(q^{2}/l)U_{0}(\alpha,\beta)$, where
\begin{eqnarray}
     U_{0}(\alpha,\beta)&=& -\frac{1}{2\alpha\beta}
     -\frac{\alpha^{2}}{\beta}\sum_{n=1}^{\infty}
     \frac{1}{n(n^{2}-\alpha^{2})}+\frac{\zeta(r)}{\beta^{r}}
     \nonumber\\
     &&+\frac{1}{2\beta^{r}}\sum_{n}\frac{1}{|n+\alpha|^{r}}
     \label{eq:U_0}
\end{eqnarray}
is the dimensionless chain energy per particle (for another
representation of $U_{0}(\alpha,\beta)$ see Appendix~A). The
equilibrium values of $\alpha$ and $\beta$ are defined by the set
of equations $\partial U_{0}(\alpha,\beta)/\partial\alpha=0$ and
$\partial U_{0}(\alpha,\beta)/\partial\beta=0$, which are written
as
\begin{eqnarray}
     \frac{1}{2\alpha^{2}\beta} &-& \frac{2\alpha}{\beta}
     \sum_{n=1}^{\infty}\frac{n}{(n^{2}-\alpha^{2})^{2}}-
     \frac{r}{2\beta^{r}}\sum_{n}\frac{n+\alpha}{|n+\alpha|^{r+2}}
     =0,\nonumber\\
     \label{eq:equilibr1}\\
     \frac{1}{2\alpha\beta^{2}} &+& \frac{\alpha^{2}}{\beta^{2}}
     \sum_{n=1}^{\infty}\frac{1}{n(n^{2}-\alpha^{2})}-\frac{r}
     {2\beta^{r+1}}\Bigg(2\zeta(r)
     \nonumber\\
     &+&\sum_{n}\frac{1}{|n+\alpha|^{r}}\Bigg)=0.
     \label{eq:equilibr2}
\end{eqnarray}
Analysis shows that these equations
have three solutions: $\alpha=\alpha_{s}(r)$ and $\beta=
\beta_{s}(r)$ $(s=1,2,3)$. Below we find $\alpha_{s}(r)$ and
$\beta_{s}(r)$, and calculate the chain equilibrium energies
$U_{s}(r)=U_{0}(\alpha_{s}(r),\beta_{s}(r))$. For illustrative
purposes we introduce also the parameters $\gamma_{s}(r)=
\alpha_{s}(r)\beta_{s}(r)$, which define the equilibrium values of
the parameter $\gamma=\alpha\beta=a/l$.

\subsection{First solution}

For $\alpha=1/2$ ($p=2a$), i.e., for a chain with equidistant
particles, the series in Eqs.~(\ref{eq:U_0}) --
(\ref{eq:equilibr2}) are either standard series or can be reduced
to them. Indeed, using standard series,\cite{PrBrMa89} we find
with linear accuracy in $\varepsilon=1/2-\alpha$
$(\varepsilon\to0)$:
\begin{eqnarray}
     \sum_{n=1}^{\infty}\frac{n}{(n^{2}-\alpha^{2})^{2}}&=&2
     -2[7\zeta(3)-6]\varepsilon,
     \label{eq:series1}\\
     \sum_{n}\frac{n+\alpha}{|n+\alpha|^{r+2}}&=&
     2(r+1)(2^{r+2}-1)\zeta(r+2)\varepsilon,\qquad
     \label{eq:series2}\\
     \sum_{n=1}^{\infty}\frac{1}{n(n^{2}-\alpha^{2})}&=&4(2\ln2-1)-
     8(3-4\ln2)\varepsilon,
     \label{eq:series3}\\
     \sum_{n}\frac{1}{|n+\alpha|^{r}}&=&
     2(2^{r}-1)\zeta(r).
     \label{eq:series4}
\end{eqnarray}
According to these formulas, $U_{0}(1/2,\beta)$ is written as
\begin{equation}
     U_{0}(1/2,\beta)=-\frac{2}{\beta}\ln2+\left(\frac{2}{\beta}
     \right)^{r}\zeta(r),
     \label{eq:U_0(a)}
\end{equation}
at $\alpha=1/2$ Eq.~(\ref{eq:equilibr1}) is satisfied identically,
i.e., $\alpha_{1}(r)=1/2$, and Eq.~(\ref{eq:equilibr2}), which at
$\alpha=1/2$ is equivalent to the equation
$dU_{0}(1/2,\beta)/d\beta=0$, yields
\begin{equation}
     \beta_{1}(r)=2\left(\frac{r\zeta(r)}{\ln2}\right)^{1/(r-1)}.
     \label{eq:beta1}
\end{equation}
Substituting $\beta=\beta_{1}(r)$ into Eq.~(\ref{eq:U_0(a)}), we
find for the equilibrium chain
\begin{equation}
     U_{1}(r)=-\ln2\,\frac{r-1}{r}\left(\frac{\ln2}{r\zeta(r)}\right)^
     {1/(r-1)}.
     \label{eq:U1}
\end{equation}
[Note that $U_{1}(r)$ is the minimum value of $U_{0}(1/2,\beta)$
only if $r>1$.]

The plots of the functions $\beta_{1}(r)$, $\gamma_{1}(r)$, and
$U_{1}(r)$ are shown in Figs.~2, 3 and 4, respectively. To gain
more insight into the behavior of these functions near $r=\infty$
and $r=1$, we define the limiting values of those functions as
$r\to\infty$ and their asymptotics as $r\to1$. Taking into account
that $\lim_{x\to0}x^{x}=1$, $\lim_{x\to0}(1+x)^{1/x}=e$,
$\zeta(\infty)=1$, and $\zeta(r)\sim (r-1)^{-1}$ as $r\to1$, we
obtain from Eq.~(\ref{eq:beta1}) $\beta_{1}(\infty)=2$ and
$\beta_{1}(r) = 2e(\epsilon\ln2)^ {-1/\epsilon}\to\infty$ as
$\epsilon=r-1\to0$, and from Eq.~(\ref{eq:U1})
$U_{1}(\infty)=-\ln2$ and $U_{1}(r) \sim -e^{-1}(\epsilon\ln2)
^{1+1/\epsilon}\to0$ as $\epsilon\to0$. Thus, if $r\to1$ then the
functions $\beta_{1}(r)$ and $U_{1}(r)$ go extremely fast to
infinity and zero, respectively. From a physical point of view,
this behavior of $\beta_{1}(r)$ and $U_{1}(r)$ is caused by the
unlimited growth of the total repulsive energy of a chain as
$r\to1$.

\subsection{Second solution}

If $\beta\neq\infty$, then we eliminate $\beta$ from
Eq.~(\ref{eq:equilibr1}) with the help of
Eq.~(\ref{eq:equilibr2}) and obtain the equation
\begin{eqnarray}
     &&\Bigg(\frac{1}{\alpha^{2}}-4\alpha\sum_{n=1}^{\infty}
     \frac{n}{(n^{2}-\alpha^{2})^{2}}\Bigg)\Bigg(2\zeta(r)+
     \sum_{n}\frac{1}{|n+\alpha|^{r}}\Bigg)
     \nonumber\\
     &&-\Bigg(\frac{1}{\alpha} + 2\alpha^{2}\sum_{n=1}^{\infty}
     \frac{1}{n(n^{2}-\alpha^{2})}\Bigg)
     \sum_{n}\frac{n+\alpha}{|n+\alpha|^{r+2}}=0,
     \nonumber\\
     \label{eq:equilibr_alpha}
\end{eqnarray}
which defines the equilibrium values of $\alpha$. According to the
results of the previous subsection, Eq.~(\ref{eq:equilibr_alpha})
always has the solution $\alpha=1/2$, i.e., $\alpha_{1}(r)=1/2$.
To determine if Eq.~(\ref{eq:equilibr_alpha}) has solutions on the
interval $(0,1/2)$, we find first the asymptotics of its left-hand
side, denoted by $L_{r}(\alpha)$, as $\alpha\to0$ and
$\alpha\to1/2$. If $\alpha\to0$, then
\begin{eqnarray}
     L_{r}(\alpha)&\sim&[\alpha^{-2} - 4\zeta(3)\alpha]
     [\alpha^{-r} + 2\zeta(r)]
     \nonumber\\
     &&-[\alpha^{-1} + 2\zeta(3) \alpha^{2}]\alpha^{-(r+1)},
     \label{eq:asympL}
\end{eqnarray}
and we obtain different asymptotics of $L_{r}(\alpha)$ for different
values of $r$:
\begin{equation}
     L_{r}(\alpha)\sim\left\{
     \begin{array}{ll}
     -6\zeta(3)\alpha^{-(r-1)},  \quad r>3, \\[3pt]
     -2\zeta(3)\alpha^{-2},  \quad r=3, \\[3pt]
     +4\zeta(r)\alpha^{-2},  \quad r<3.
     \end{array}
     \right.
     \label{eq:asympL1}
\end{equation}

Using Eqs.~(\ref{eq:series1})--(\ref{eq:series4}), as
$\alpha\to1/2$ we obtain
\begin{equation}
     L_{r}(\alpha) \sim -2^{r+3}\zeta(r)F(r)\varepsilon,
     \label{eq:asympL2}
\end{equation}
where
\begin{equation}
     F(r)=4(1-2^{-r-2})(r+1)\ln2\frac{\zeta(r+2)}{\zeta(r)}
     -7\zeta(3)
     \label{eq:defF}
\end{equation}
is a monotone increasing function of $r$, and $F(r)=0$ at
$r=r_{1}\approx2.779$. Therefore, in the left vicinity of the
point $\alpha=1/2$, the conditions $L_{r}(\alpha)>0$ and
$L_{r}(\alpha)<0$ hold for $r<r_{1}$ and $r>r_{1}$, respectively.
Using this fact and Eq.~(\ref{eq:asympL1}), we conclude that if
$r_{1}<r<3$, then $L_{r}(\alpha)$ near the points $\alpha=0$ and
$\alpha=1/2$ has different signs, and so
Eq.~(\ref{eq:equilibr_alpha}) has at last one solution
$\alpha_{2}(r)\in(0,1/2)$ satisfying the conditions $\alpha_{2}(r)
\to 1/2$ as $r\to r_{1}$ and $\alpha_{2}(r)\to0$ as $r\to 3$.
Numerical analysis shows that such a solution is unique, and
Eq.~(\ref{eq:equilibr2}) yields
\begin{equation}
     \beta_{2}(r)=\left[r\frac{2\zeta(r)+\displaystyle\sum_
     {n}\frac{1}{|n+\alpha_{2}(r)|^{r}}}
     {\displaystyle\frac{1}{\alpha_{2}(r)}+\displaystyle
     2\alpha_{2}^{2}(r)\sum_{n=1}^{\infty}\frac{1}{n[n^{2}-
     \alpha_{2}^{2}(r)]}}\right]^{\frac{1}{r-1}}.
     \label{eq:beta2}
\end{equation}

Since  a chain is transformed to a chain with equidistant
particles as $r\to r_{1}$, the equalities $\beta_{2}(r_{1}) =
\beta_{1}(r_{1})$, $\gamma_{2}(r_{1}) = \gamma_{1}(r_{1})$, and
$U_{2}(r_{1}) = U_{1}(r_{1})$ hold. If $r\to3$, then according to
Eq.~(\ref{eq:beta2}) $\beta_{2}(r)\sim \sqrt{3} /\alpha_{2}(r) \to
\infty$, and we obtain $\gamma_{2}(3) = \sqrt{3}$,
$U_{2}(3)=-1/3\sqrt{3}$. The plots of the functions
$\beta_{2}(r)$, $\gamma_{2}(r)$, and $U_{2}(r)$, which are defined
for $r\in[r_{1},3]$, are also shown in Figs.~2--4. As is seen from
Fig.~4, $U_{2}(r)>U_{1}(r)$ for $r_{1} < r \leq 3$, i.e., the
equilibrium chain with equidistant particles is energy-wise
preferred compared to the equilibrium chain with non-equidistant
particles.

\subsection{Third solution}

A further solution of Eqs.~(\ref{eq:equilibr1}) and
(\ref{eq:equilibr2}) corresponds to the case when the period of
the chain is infinite and the distance between particles in the
cells is finite, i.e., when $\alpha\to0$, $\beta\to\infty$, and
$\alpha\beta=\gamma$. In this case
\begin{equation}
     U_{0}(\alpha,\beta)\equiv U_{0}(\gamma)=-\frac{1}{2\gamma}
     +\frac{1}{2\gamma^{r}},
     \label{eq:U_03}
\end{equation}
Eq.~(\ref{eq:equilibr2}) is satisfied identically, and
Eq.~(\ref{eq:equilibr1}) is reduced to the equation
$dU_{0}(\gamma)/d\gamma =0$. Its solution $\gamma=\gamma_{3}(r)$,
$\gamma_{3}(r)= r^{1/(r-1)}$, defines the equilibrium distance
between the particles, and the equilibrium energy per particle
is given by
\begin{equation}
     U_{3}(r)=-\frac{r-1}{2r}\left(\frac{1}{r}\right)^{1/(r-1)}.
     \label{eq:U3}
\end{equation}
The functions $\gamma_{3}(r)$ and $U_{3}(r)$ monotonically
decrease as $r$ grows from $1$ to $\infty$ [$U_{3}\sim(r-1)
/2e\to0$ as $r\to1$, $U_{3}(3)=-1/3\sqrt{3}$, $U_{3}(\infty)
=-1/2$, $\gamma_{3}(1)=e$, $\gamma_{3}(3)= \sqrt{3}$, $\gamma_{3}
(\infty)=1$], and their plots are shown in Figs.~3 and 4.
Comparing the functions $U_{1}(r)$, $U_{2}(r)$, and $U_{3}(r)$, we
conclude that for $r>r_{2}$ [$r=r_{2}\approx2.799$ is the solution
of the equation $U_{1}(r)=U_{3}(r)$] the minimum energy occurs for
a chain with equidistant particles [$p=2a$, $a=l\gamma_{1}(r)$],
and for $r<r_{2}$ the minimum energy occurs for a chain with
infinite period [$p=\infty$, $a=l\gamma_{3}(r)$]. Note also that a
chain with non-equidistant particles has a slightly larger energy
than other chains.

Next we study the vibration properties of a chain with equidistant
particles.

\section{Vibration properties}

\subsection{Dispersion relation}

Assuming that the condition $|x_{n}^{\pm}|\ll a$ holds for all
particles, we expand the  chain energy $W_{int}$ in a power series
of $x_{n}^{\pm}/a$ with quadratic accuracy: $W_{int}=W_{int}^
{(0)}+W_{int}^{(1)}+W_{int}^{(2)}$, where the upper indices
indicate the power of $x_{n}^{\pm}$. Straightforward calculations
based on Eq.~(\ref{eq:energy_gen2}) with $p=2a$ and arbitrary
value of $a$ yield $W_{int}^{(0)}=N(q^{2}/l)U_{0}(1/2,2a/l)$,
$W_{int}^{(1)}=0$, and
\begin{eqnarray}
     W_{int}^{(2)}&=&\frac{1}{4}{\sum_{n,m}}'B_{2(n-m)}
     \left[(x_{n}^{+}-x_{m}^{+})^{2}+(x_{n}^{-}-
     x_{m}^{-})^{2}\right]
     \nonumber\\
     &&+\frac{1}{2}\sum_{n,m}B_{2(n-m)-1}(x_{n}^{+}-
     x_{m}^{-})^{2},
     \label{eq:defW2}
\end{eqnarray}
where
\begin{equation}
     B_{n}=\frac{(-1)^{n}2q^{2}}{a^{3}|n|^{3}}+\frac{br(r+1)}
     {a^{r+2}|n|^{r+2}}.
     \label{eq:defB}
\end{equation}
For the case considered here that $a$ has an equilibrium value,
i.e., $a=l\gamma_{1}(r)$, we represent $B_{n}$ in the form $B_{n}=
M\Omega^{2}\tilde{B}_{n}$, where $\Omega^{2}=q^{2}/Ml^{3}$,
\begin{equation}
     \tilde{B}_{n}=\frac{1}{\gamma_{1}^{3}(r)}\left[2
     \frac{(-1)^{n}}{|n|^{3}}+\frac{(r+1)\ln2}{\zeta(r)
     |n|^{r+2}}\right],
     \label{eq:def_tildeB}
\end{equation}
and according to Eq.~(\ref{eq:beta1}) $\gamma_{1}(r)=
[r\zeta(r)/\ln2]^{1/(r-1)}$.

 From Eq.~(\ref{eq:defW2}), we obtain the system of equations,
$M\ddot{x}_{n}^{\pm}+\partial W_{int}^{(2)}/\partial
x_{n}^{\pm}=0$, that describe free longitudinal vibrations of
particles near their equilibrium positions in the harmonic
approximation, in the form
\begin{equation}
     \begin{array}{ll}
     \displaystyle \ddot{x}_{n}^{+}+\Omega^{2}\sum_{m}
     \tilde{B}_{2(n-m)-1}(x_{n}^{+}-x_{m}^{-})
     \\[14pt]
     \displaystyle \phantom{\ddot{x}_{n}^{+}}+\Omega^{2}\sum_
     {m\neq n}\tilde{B}_{2(n-m)}(x_{n}^{+}-x_{m}^{+})=0,
     \\[16pt]
     \displaystyle \ddot{x}_{n}^{-}+\Omega^{2}\sum_{m}\tilde{B}_
     {2(n-m)+1}(x_{n}^{-}-x_{m}^{+})
     \\[14pt]
     \displaystyle \phantom{\ddot{x}_{n}^{-}}+\Omega^{2}\sum_
     {m\neq n}\tilde{B}_{2(n-m)}(x_{n}^{-}-x_{m}^{-})=0.
     \end{array}
     \label{eq:eq_pm}
\end{equation}
Since $n$ ranges over the integers, this system contains infinitely
many coupled equations. We seek wave solutions, which in the
complex representation have the form
\begin{equation}
     x_{n}^{\pm}=x_{\pm}\exp(i2nak-i\omega t).
     \label{eq:comlex}
\end{equation}
Here $x_{+}$ and $x_{-}$ are the complex amplitudes, $k$ is the
wave number belonging to the Brillouin zone, i.e., to the region
$[-\pi/2a,\pi/2a]$ of reciprocal space, and $\omega$ is the wave
frequency. Substituting solution (\ref{eq:comlex}) into
Eqs.~(\ref{eq:eq_pm}) and using the definition $\kappa=ak/\pi$
$(\kappa\in[-1/2,1/2])$,
\begin{eqnarray}
     D&=&2\sum_{p=1}^{\infty}\tilde{B}_{p}\,,
     \label{eq:defD}\\
     P(\kappa)&=&2\sum_{p=1}^{\infty}\tilde{B}_{2p}\cos(2p\pi
     \kappa),
     \label{eq:defP}\\
     S(\kappa)&=&\sum_{p}\tilde{B}_{2p-1}\exp(-i2p\pi\kappa),
     \label{eq:defS}
\end{eqnarray}
we obtain the system of equations for $x_{+}$ and $x_{-}$
\begin{equation}
     \begin{array}{ll}
     \phantom{-*}x_{+}[-\tilde{\omega}^{2}+D-P(\kappa)]-x_{-}
     S(\kappa)=0,
     \\[8pt]
     -x_{+}S^{*}(\kappa)+x_{-}[-\tilde{\omega}^{2}+D-P(\kappa)]=0,
     \end{array}
     \label{eq:eq_ampl}
\end{equation}
where $\tilde{\omega}=\omega/\Omega$ is the dimensionless wave
frequency, and the asterisk denotes complex conjugation. This
system has non-trivial solutions, i.e, propagation of phonons
along a chain can occur, if its determinant equals
zero,\cite{Lankaster}
\begin{equation}
     [\tilde{\omega}^{2}-D+P(\kappa)]^{2}-|S(\kappa)|^{2}=0.
     \label{eq:determinant}
\end{equation}
Solving Eq.~(\ref{eq:determinant}) with respect to
$\tilde{\omega}$, we find the dispersion relation for phonons:
$\omega_{1,2}(\kappa)=\Omega\tilde{\omega}_ {1,2}(\kappa)$, where
\begin{equation}
     \tilde{\omega}_{1,2}(\kappa)=\sqrt{D-P(\kappa)
     \pm |S(\kappa)|},
     \label{eq:spectr}
\end{equation}
and the plus and minus sign in front of $|S(\kappa)|$ correspond to
$\tilde{\omega}_{1} (\kappa)$ and $\tilde{\omega}_{2} (\kappa)$,
respectively.

\subsection{Analysis of the spectrum}

Since $\tilde{\omega}_{1,2} (-\kappa)=\tilde{\omega}_{1,2}
(\kappa)$, we can restrict the permissible values of the
dimensionless wave number $\kappa$ by the condition $\kappa \in
[0,1/2]$. According to Eq.~(\ref{eq:spectr}), in this interval the
dimensionless phonon spectrum consists of two branches,
$\tilde{\omega}_{1} (\kappa)$ and $\tilde{\omega}_{2} (\kappa)$.
It is easy to see that on the interval $[0,1]$ the spectrum is
given by
\begin{equation}
     \tilde{\omega}(\kappa)=\left\{
     \begin{array}{ll}
     \tilde{\omega}_{1}(\kappa),  \quad 0\leq\kappa\leq1/2, \\[3pt]
     \tilde{\omega}_{2}(1-\kappa),  \quad 1/2<\kappa\leq1,
     \end{array}
     \right.
     \label{eq:spectr_ext}
\end{equation}
and $\tilde{\omega}(\kappa)$ is continuous at $\kappa= 1/2$.
Indeed, for $\kappa=1/2$ Eq.~(\ref{eq:defS}) yields
$S(1/2)=\sum_{p}(-1)^{p} \tilde{B}_{2p-1}$. Replacing in this
formula $p$ by $-p+1$ and using the condition $\tilde{B}_{n}
=\tilde{B}_{-n}$, we obtain $S(1/2)=-\sum_{p}(-1)^{p}
\tilde{B}_{2p-1}$. This implies that $S(1/2)=0$, and
Eq.~(\ref{eq:spectr}) yields $\tilde{\omega}_{1}(1/2)=
\tilde{\omega}_{2}(1/2)$. Hence, the phonon spectrum on the
interval $[0,1]$ is represented by the single branch
$\tilde{\omega}(\kappa)$. Note that the region $[-1,1]$ is the
dimensionless Brillouin zone corresponding to a chain with a
primitive cell which contains one particle.

To study the behavior of the phonon spectrum $\tilde{\omega}
(\kappa)$, we use Eq.~(\ref{eq:def_tildeB}) to
represent Eqs.~(\ref{eq:defD})--(\ref{eq:defS}) in the form
\begin{eqnarray}
     D&=&\frac{1}{\gamma_{1}^{3}(r)}\bigg[2(r+1)\ln2\frac{
     \zeta(r+2)}{\zeta(r)}-3\zeta(3)\bigg],
     \label{eq:D}\\
     P(\kappa)&=&\frac{1}{2\gamma_{1}^{3}(r)}\bigg[
     \frac{(r+1)\ln2}{2^{r}\zeta(r)}\Phi_{r+2}
     (\kappa)+\Phi_{3}(\kappa)\bigg],
     \label{eq:P}\\
     S(\kappa)&=&\frac{4e^{-i\pi\kappa}}{\gamma_{1}^{3}(r)}
     \bigg[\frac{(r+1)\ln2}{2\zeta(r)}\Psi_{r+2}(\kappa)
     -\Psi_{3}(\kappa)\bigg],\qquad
     \label{eq:S}
\end{eqnarray}
where
\begin{eqnarray}
     \Phi_{s}(\kappa)&=&\sum_{p=1}^{\infty}\frac{\cos(2p\pi
     \kappa)}{p^{s}},
     \label{eq:defPhi}\\
     \Psi_{s}(\kappa)&=&\sum_{p=1}^{\infty}\frac{\cos[(2p-1)
     \pi\kappa]}{(2p-1)^{s}}.
     \label{eq:defPsi}
\end{eqnarray}
[Recall that the functions $P(\kappa)$ and $S(\kappa)$ are defined
for $|\kappa|\leq1/2$.] Unfortunately, the analytical summation of
the series in Eqs.~(\ref{eq:P}) and (\ref{eq:S}) can be performed
only for special cases, such as $\kappa=0$ or $1/2$. We calculate
$\tilde{\omega}(\kappa)$ only in the centre $(\kappa=0)$ and on
the boundary $(\kappa=1)$ of the Brillouin zone which, according
to Eq.~(\ref{eq:spectr_ext}), requires us to determine $P(0)$ and
$S(0)$. Using the definition of the Riemann zeta function and the
standard series $\sum_{p=1}^{\infty} (2p-1)^{-s}=
(1-2^{-s})\zeta(s)$, we obtain
\begin{equation}
     P(0)=\frac{1}{2\gamma_{1}^{3}(r)}\bigg[2^{-r}(r+1)\ln2\frac{
     \zeta(r+2)}{\zeta(r)}+\zeta(3)\bigg]
     \label{eq:P(0)}
\end{equation}
and $D-P(0)=S(0)=F(r)/2\gamma_{1}^{3}(r)$, where $F(r)$ is defined
by Eq.~(\ref{eq:defF}). If $F(r)>0$, i.e., $r>r_{1}$, then
\begin{equation}
     \tilde{\omega}(0)=0, \qquad
     \tilde{\omega}(1)=\sqrt{F(r)/\gamma_{1}^{3}(r)}\,.
     \label{eq:omega_0_1}
\end{equation}
Note that Eqs.~(\ref{eq:eq_ampl}) yield $x_{+}=x_{-}$ for
$\tilde{\omega}=0$, and $x_{+}=-x_{-}$ for $\tilde{\omega}=
\tilde{\omega}(1)$. This means that neighboring particles in a
chain perform acoustical vibrations at $\kappa=0$ and optical ones at
$\kappa=1$.

To find for $r>r_{1}$ the asymptotics of $\tilde{\omega} (\kappa)$
as $\kappa\to0$ and $\kappa\to1$, we need to calculate the series
(\ref{eq:defPhi}) and (\ref{eq:defPsi}) with quadratic accuracy in
$\kappa$ as $\kappa\to0$. This is not a problem for the series
$\Phi_{r+2}(\kappa)$ and $\Psi_{r+2}(\kappa)$ that arise from the
repulsive interaction of particles, because expanding the cosines in a
power series of $\kappa$ to an accuracy of $\kappa^{2}$ we
immediately obtain
\begin{eqnarray}
     \Phi_{r+2}(\kappa)&=&\zeta(r+2)-2\pi^{2}\zeta(r)\kappa^{2},
     \label{eq:Phi_r+2}\\
     \Psi_{r+2}(\kappa)&=&(1-2^{-r-2})\zeta(r+2)-\frac{\pi^{2}}{2}
     (1-2^{-r})\zeta(r)\kappa^{2}.\nonumber\\
     \label{eq:Psi_r+2}
\end{eqnarray}
As to the series $\Phi_{3}(\kappa)$ and $\Psi_{3}(\kappa)$ that
arise from the Coulomb interaction, their calculation by this
method is impossible since divergencies occur. To avoid this
difficulty, we use their integral representations obtained in
Appendix~B. These representations have the advantage that the
double integrals in Eqs.~(\ref{eq:intPhi}) and (\ref{eq:intPsi})
can be easily calculated with arbitrary accuracy as $\kappa\to0$.
Specifically, if we use the approximate formulas $\ln[2\sin (y/2)]
\approx \ln y$ and $\ln[\tan(y/2)] \approx \ln(y/2)$ $(y \ll 1)$,
then Eqs.~(\ref{eq:intPhi}) and (\ref{eq:intPsi}) yield
\begin{eqnarray}
     \Phi_{3}(\kappa)&=&\zeta(3)-\pi^{2}[3-2\ln(2\pi\kappa)]
     \kappa^{2},
     \label{eq:Phi_3}\\
     \Psi_{3}(\kappa)&=&\frac{7}{8}\zeta(3)-\frac{\pi^{2}}{8}
     \left(3-2\ln\frac{\pi\kappa}{2}\right)\kappa^{2}.
     \label{eq:Psi_3}
\end{eqnarray}

Substituting series (\ref{eq:Phi_r+2})--(\ref{eq:Psi_3}) into
Eqs.~(\ref{eq:P}) and (\ref{eq:S}), we find the main terms of the
asymptotic expansion of $\tilde{\omega}(\kappa)$ as $\kappa\to0$,
\begin{equation}
     \tilde{\omega}(\kappa) \sim \pi\sqrt{\frac{(r-1)\ln2}
     {\gamma_{1}^{3}(r)}}\,\kappa,
     \label{eq:asymp1}
\end{equation}
and as $\kappa\to1$,
\begin{equation}
     \tilde{\omega}(\kappa) \sim \tilde{\omega}(1) + \frac
     {\pi^{2}}{\tilde{\omega}(1)\gamma_{1}^{3}(r)}
     (1-\kappa)^{2}|\ln(1-\kappa)|.
     \label{eq:asymp2}
\end{equation}
According to these asymptotics, the long-wave part of the
vibrational spectrum (for $\kappa\ll1$) has the usual form, i.e.,
$\tilde{\omega}(\kappa) \propto \kappa$. On the boundary of the
Brillouin zone, the function $\tilde{\omega} (\kappa)$ has a local
minimum, $\tilde{\omega}'(1)=d\tilde{\omega} (\kappa)
/d\kappa|_{\kappa=1}=0$, and the short-wave part of the spectrum
(for $\kappa\sim1$) exhibits a non-analytic dependence on
$\kappa$. As Eqs.~(\ref{eq:Phi_3}) and (\ref{eq:Psi_3}) show, the
non-analyticity of $\tilde{\omega}(\kappa)$ at $\kappa\to1$ is
caused by the long-range action of the Coulomb interaction.

The dependence of $\tilde{\omega}(\kappa)$ on $\kappa$ for
different values of $r$ is shown in Fig.~5. As is seen from this
figure, $\tilde{\omega}(\kappa)$ is a non-monotonic function of
$\kappa$ which has a unique maximum on the interval $(1/2,1)$. As
$r$ decreases, its local minimum $\tilde{\omega}(1)$ also
decreases, and $\tilde{\omega}(1)>0$ for $r>r_{1}$. Since the
inequality $\tilde{\omega}(1)>0$ implies that
$\tilde{\omega}(\kappa)>0$ for all $\kappa\neq0$, the condition
$r>r_{1}$ is the stability criterion of a chain with respect to
small displacements of particles from their equilibrium positions.
Note that though for $r_{1}<r<r_{2}$ a chain with infinite period
is energy-wise preferred compared to a chain with equidistant
particles, it is the latter which is stable. If $r<r_{1}$, then a
region exists on the interval $[0,1]$ where
$\tilde{\omega}^{2}(\kappa)<0$, and therefore such chains are
unstable. Finally, if $r=r_{1}$, then $\tilde{\omega}(1)=0$,
$\tilde{\omega}'(1)=-\infty$, and the asymptotic of
$\tilde{\omega}(\kappa)$ as $\kappa\to1$ has the form
\begin{equation}
     \tilde{\omega}(\kappa) \sim \pi\sqrt{\frac{2}{\gamma_{1}^{3}
     (r_{1})}}\,(1-\kappa)\sqrt{|\ln(1-\kappa)|}.
     \label{eq:asymp3}
\end{equation}
In this case the optical displacements of particles do not lead to
an appearance of restoring forces. Therefore, to determine if a
chain is stable or not at $r=r_{1}$, it is necessary to go beyond
the harmonic approximation.

\subsection{Forced vibrations}

We briefly consider also the forced vibrations of particles under
the action of the longitudinal alternating electric field
$E(t)=E\cos\omega_{0}t$, where $E$ and $\omega_{0}$ are the
amplitude and frequency of the alternating field, respectively. In
such field the electric force $qE\cos\omega _{0}t$ acts on
positive charges, and the force $-qE\cos\omega_{0}t$ acts on
negative charges. If the friction force $-\lambda\dot{x}_
{n}^{\pm}$ ($\lambda$ is a damping coefficient) also acts on each
particle, then the forced vibrations of particles are described by
the following system of motion equations:
\begin{equation}
     \begin{array}{ll}
     \displaystyle \ddot{x}_{n}^{+}+2\Omega_{\lambda}\dot{x}_{n}^{+}+
     \Omega^{2}\sum_{m}\tilde{B}_{2(n-m)-1}(x_{n}^{+}-x_{m}^{-})
     \\[14pt]
     \displaystyle \phantom{\ddot{x}_{n}^{+}}+\Omega^{2}\sum_{m\neq
     n}\tilde{B}_{2(n-m)}(x_{n}^{+}-x_{m}^{+})=A\cos\omega_{0}t,
     \\[16pt]
     \displaystyle \ddot{x}_{n}^{-}+2\Omega_{\lambda}\dot{x}_{n}^{-}+
     \Omega^{2}\sum_{m}\tilde{B}_{2(n-m)+1}(x_{n}^{-}-x_{m}^{+})
     \\[14pt]
     \displaystyle \phantom{\ddot{x}_{n}^{-}}+\Omega^{2}\sum_{m\neq n}
     \tilde{B}_{2(n-m)}(x_{n}^{-}-x_{m}^{-})=-A\cos\omega_{0}t
     \end{array}
     \label{eq:eq_forced}
\end{equation}
$(\Omega_{\lambda}=\lambda/2M$, $A=qE/M)$. Symmetry of these
equations suggests that we seek their solution in the form
$x_{n}^{+}=x(t)$ and $x_{n}^{-}=-x(t)$. Taking into account that
$\tilde{\omega}^{2}(1)=2\sum_{p=1}^{\infty} \tilde{B}_{2p-1}$, in
this case Eqs.~(\ref{eq:eq_forced}) are reduced to the ordinary
differential equation for a forced harmonic oscillator
\begin{equation}
     \ddot{x}(t)+2\Omega_{\lambda}\dot{x}(t)+\omega^{2}(1)
     x(t)=A\cos \omega_{0}t.
     \label{eq:eq_delta}
\end{equation}

Its steady-state solution, which describes the forced vibrations
of particles, has the form $x(t)=x_{0}\cos(\omega_{0}t-\varphi)$,
where
\begin{equation}
     x_{0}=\frac{A}{\displaystyle\sqrt{[\omega^{2}(1)-
     \omega_{0}^{2}]^{2}+4\omega_{0}^{2}\Omega_{\lambda}^{2}}}
     \label{eq:delta}
\end{equation}
is the amplitude displacement of particles from their equilibrium
positions, and $\tan\varphi=2\Omega_{\lambda}\omega/
[\omega^{2}(1)- \omega_{0}^{2}]$. According to
Eq.~(\ref{eq:delta}), if $\omega(1)>\sqrt{2} \Omega_{\lambda}$,
then the dependence of $x_{0}$ on the driving frequency
$\omega_{0}$ has resonance character; $\omega_{0}= [\omega^{2}(1)-
2\Omega_{\lambda}^{2}]^{1/2}$ is the resonance frequency, and
$x_{0}=A/2\Omega_{\lambda} [\omega^{2}(1)- \Omega_{\lambda}^{2}]
^{1/2}$ is the resonance amplitude.

\section{Deterministic transport}

In this section we study the transport properties of a chain whose
particles are driven by the longitudinal alternating electric
field $E(t)=Eh(t)$,
\begin{equation}
     h(t)=\left\{
     \begin{array}{ll}
     1,  \quad 2k\tau< t \leq 2k\tau + \tau, \\[3pt]
     -1, \quad 2k\tau + \tau < t\leq 2k\tau + 2\tau,
     \end{array}
     \right.
     \label{eq:el_field}
\end{equation}
and interact with the sawtooth (ratchet) potential $V(x)$ that
induces the force field $f(x)=-dV(x)/dx\equiv f_{0}g(x)$, where
\begin{equation}
     g(x)=\left\{
     \begin{array}{ll}
     c/(d-c), \quad md+c<x\leq md+d, \\[3pt]
     -1,  \quad md<x\leq md+c.
     \end{array}
     \right.
     \label{eq:ratchet_field}
\end{equation}
Here $k$ and $m$ run over all integers, $\tau$ is the half-period
of $E(t)$, $f_{0}=|\min{f(x)|}$, and $d$ is the period of $V(x)$
(see Fig.~6). The dynamics of particles in such a chain is
described by the system of equations (\ref{eq:eq_forced}) in which
the right-hand sides $A\cos\omega_{0}t$ and $-A\cos\omega_{0}t$
must be replaced by $Ah(t)+Rg(x_{n}^{+})$ and $-Ah(t)+
Rg(x_{n}^{-})$ ($R=f_{0}/M$), respectively. To simplify the
problem, we assume that the condition $a=Ld$ ($L$ is a natural
number) holds, so that the equilibrium positions of all particles
coincide with the minima of $V(x)$. In this case, under the action
of the electric field $E(t)$ all positively charged particles and
all negatively charged particles move identically, i.e.,
$x_{n}^{+}=w(t)$ and $x_{n}^{-}=u(t)$, and we obtain
\begin{equation}
     \begin{array}{ll}
     \ddot{w} + 2\Omega_{\lambda}\dot{w}+\omega^{2}(1)(w-u)
     =Ah(t)+Rg(w),
     \\[6pt]
     \ddot{u} + 2\Omega_{\lambda}\dot{u}+\omega^{2}(1)(u-w)
     =-Ah(t)+Rg(u).
     \end{array}
     \label{eq:eq_ratch}
\end{equation}

Due to the periodicity of $h(t)$ and $g(x)$, the steady-state
solution of Eqs.~(\ref{eq:eq_ratch}) satisfies the conditions
\begin{equation}
     w(t+\tau)=u(t)+Kd, \quad u(t+\tau)=w(t)+Kd,
     \label{eq:symmetry}
\end{equation}
where $K$ is a whole number. It follows from these conditions that
\begin{equation}
     w(t+2\tau)-w(t)=2Kd, \quad u(t+2\tau)-u(t)=2Kd,
     \label{eq:conditions}
\end{equation}
i.e., for the period of $h(t)$ all particles are displaced  the
same distance $\Delta=2Kd$. This fact permits us to define the
average velocity of a chain as
\begin{equation}
     v\equiv\frac{w(2\tau)-w(0)}{2\tau}=K\frac{d}{\tau}\,.
     \label{eq:define_V}
\end{equation}

To find $v$, we need to calculate $\Delta$ and in the general case
solve Eqs.~(\ref{eq:eq_ratch}). These equations have the important
feature that we need to consider only the ratchet potentials with
$c> d/2$. Indeed, the transformation $c\to d-c$ leads to the
inverted potential $V_{in}(x)= V(-x)$ which gives rise to the
reduced force field $g_{in}(x)= -g(-x)$. According to
Eqs.~(\ref{eq:eq_ratch}), in such a potential the displacements
$w_{in}(t)$ and $u_{in}(t)$ of positive and negative charges from
their equilibrium positions are governed by the equations of
motion
\begin{equation}
     \begin{array}{ll}
     \ddot{w}_{in}\! + 2\Omega_{\lambda}\dot{w}_{in}\!+
     \omega^{2}(1)(w_{in}\!-u_{in})=Ah(t)-Rg(-w_{in}),
     \\[6pt]
     \ddot{u}_{in}\! + 2\Omega_{\lambda}\dot{u}_{in}\!+
     \omega^{2}(1)(u_{in}\!-w_{in})=\!-Ah(t)-Rg(-u_{in}).\\[3pt]
     \end{array}
     \label{eq:eq_invert}\\[3pt]
\end{equation}
Comparing these equations with Eq.~(\ref{eq:eq_ratch}), we obtain
$w_{in}(t)=-u(t)$ and $u_{in}(t)=-w(t)$. This means that if for
$c> d/2$ the solution of equations (\ref{eq:eq_ratch}) is known,
then their solution for $c<d/2$ is known too. Using
Eqs.~(\ref{eq:conditions}) and (\ref{eq:define_V}), we find the
chain velocity in the inverted potential $v_{in}=-v$, i.e., in the
potentials $V_{in}(x)$ and $V(x)$ a chain drifts in the opposite
directions with the same velocity. If $c=d/2$ then $V_{in}(x)=
V(x)$, both conditions $v_{in}=-v$ and $v_{in}=v$ hold, and
  $v_{in}=v=0$. Hence, to study the transport properties of a
chain, we indeed can consider only those ratchet potentials for
which $c>d/2$. In this case, a chain can drift only along the
positive direction of the $x$ axis and so $K\geq0$.

Next we study the deterministic transport of a chain in the
overdamped limit where Eqs.~(\ref{eq:eq_ratch}) are reduced to
\begin{equation}
     \begin{array}{ll}
     2\Omega_{\lambda}\dot{w}+\omega^{2}(1)(w-u)
     =Ah(t)+Rg(w),
     \\[6pt]
     2\Omega_{\lambda}\dot{u}+\omega^{2}(1)(u-w)
     =-Ah(t)+Rg(u).
     \end{array}
     \label{eq:eq_overdamped}
\end{equation}
Since our aim is to find the chain velocity $v$, we need to solve
Eqs.~(\ref{eq:eq_overdamped}) only for $t\in(0,2\tau)$. To
simplify the problem further, we assume that the conditions $L\gg
K\gg 1$ and $1<A/R<c/(d-c)$ hold. The former implies that the
interparticle distance $a$ is much larger than the chain
displacement $\Delta$, which, in turn, is much larger than the
potential period $d$. The latter shows that for $h(t)>0$
($h(t)<0$) positive (negative) charges can overcome the potential
barriers of $V(x)$ and they move along the positive direction of
the $x$ axis, whereas negative (positive) charges stay in the
minima of $V(x)$. After $h(t)$ changes sign, the particles at rest
begin to move, and the moving particles move to the appropriate
minima of $V(x)$ and stop there. Since the maximal distance that
the moving particles travel after $h(t)$ changes sign does not
exceed $c$, i.e., this distance is much less than $\Delta$, we
assume that the moving particles reach the corresponding minima of
$V(x)$ immediately as $h(t)$ changes sign. Within this
approximation, the dynamics of positive and negative charges is
described separately, and Eqs.~(\ref{eq:eq_overdamped}) yield
$u(t)=0$ and
\begin{equation}
     2\Omega_{\lambda}\dot{w}+\omega^{2}(1)w =A+Rg(w)
     \label{eq:eq_positive}
\end{equation}
[$w(0)=-\Delta/2$, $w(\tau)=\Delta/2$] for $t\in(0,\tau)$, and
$w(t)=\Delta/2$ and
\begin{equation}
     2\Omega_{\lambda}\dot{u}+\omega^{2}(1)(u-\Delta/2)
     =A+Rg(u)
     \label{eq:eq_negative}
\end{equation}
[$u(\tau)=0$, $u(2\tau)=\Delta$] for $t\in(\tau,2\tau)$.

To calculate $\Delta$ we can use either Eq.~(\ref{eq:eq_positive})
or Eq.~(\ref{eq:eq_negative}). We take Eq.~(\ref{eq:eq_positive})
as our starting point and write it in the form
\begin{equation}
     \dot{w}+\delta w =\left\{
     \begin{array}{ll}
     \theta_{2}, \quad md+c<w\leq md+d, \\[3pt]
     \theta_{1},  \quad md<w\leq md+c,
     \end{array}
     \right.
     \label{eq:eq_basic}
\end{equation}
where $\delta=\omega^{2}(1)/2\Omega_{\lambda}$, $\theta_{1}=
(A-R)/2\Omega_{\lambda}$, $\theta_{2}= [A+Rc/(d-c)]/
2\Omega_{\lambda}$, and $m=-K,-K+1,...,K-1$. The solution of
Eq.~(\ref{eq:eq_basic}) depends on $\Delta$ and actually defines
$\Delta$, because $w(0)=-\Delta/2$ and $w(\tau)=\Delta/2$. We use
this feature to derive the following equation (see Appendix~C)
\begin{eqnarray}
     e^{\delta t_{2K}}&=&\frac{\displaystyle\Gamma\left(\frac
     {2\theta_{1}+\Delta\delta}{2d\delta}+1\right)}
     {\displaystyle\Gamma\left(\frac{2\theta_{1}-\Delta\delta}
     {2d\delta}+1\right)}\,\frac{\displaystyle\Gamma\left(\frac
     {2\theta_{1}-\Delta\delta}{2d\delta}-\frac{c}{d}+1\right)}
     {\displaystyle\Gamma\left(\frac{2\theta_{1}+\Delta\delta}
     {2d\delta}-\frac{c}{d}+1\right)}
     \nonumber\\[3pt]
     &&\times\frac{\displaystyle\Gamma\left(\frac{2\theta_{2}+
     \Delta\delta}{2d\delta}-\frac{c}{d}+1\right)}{\displaystyle
     \Gamma\left(\frac{2\theta_{2}-\Delta\delta}{2d\delta}-
     \frac{c}{d}+1\right)}\,\frac{\displaystyle\Gamma\left(\frac
     {2\theta_{2}-\Delta\delta}{2d\delta}\right)}{\displaystyle
     \Gamma\left(\frac{2\theta_{2}+\Delta\delta}{2d\delta}\right)}
     \nonumber\\
     \label{eq:eq_Delta}
\end{eqnarray}
for the instant $t=t_{2K}$ that corresponds to a displacement of
positive charges of $\Delta=2Kd$. An increase of $\tau$ leads to
the step-wise increase of $t_{2K}$ and $K$, and from
Eq.~(\ref{eq:eq_Delta}) we obtain
\begin{equation}
     \max{K}=\min{\left\{\left[\frac{\theta_{1}+(d-c)\delta}
     {d\delta}\right],\left[\frac{\theta_{2}}{d\delta}\right]
     \right\}},
     \label{eq:def_maxK}
\end{equation}
where $[x]$ denotes the integer part of $x$.

Replacing in Eq.~(\ref{eq:eq_Delta}) $t_{2K}$ by $\tau$, we obtain
the approximate equation that defines $\Delta$ as a function of
$\tau$. For $\delta\tau\gg1$ it yields $K\approx\max{K}$, and so
in this case the chain velocity $v$ is given by
\begin{equation}
     v=\frac{2d}{\tau}\max{K}.
     \label{eq:velocity1}
\end{equation}
It follows from this formula that $v\to0$ as $\tau\to\infty$ and,
according to Eq.~(\ref{eq:def_maxK}), $v$ depends weakly on $d$
and $v\propto\omega^{-2}(1)$. The last proportionality shows that
the chain velocity strongly depends on the interparticle
interactions, and $v$ is a decreasing function of $r$. [Note that
the permissible values of $r$ are not too close to $r_{1}$, since
our assumptions imply that $\max{K}\ll L$ holds.]

If $K\ll \max{K}$, then, using the asymptotic formula\cite{BaEr53}
\begin{equation}
     \frac{\Gamma(z+x)}{\Gamma(z+y)}=z^{x-y}\left[1+\frac{1}{2z}
     (x-y)(x+y-1)\right]
     \label{eq:asympt1}
\end{equation}
($z\to\infty$), we reduce the equation for $\Delta$ to the form
\begin{equation}
     e^{\delta\tau}=1+\frac{\Delta\delta}{\theta_{1}}\frac{c}{d}+
     \frac{\Delta\delta}{\theta_{2}}\left(1-\frac{c}{d}\right).
     \label{eq:eq_Delta1}
\end{equation}
Solving Eq.~(\ref{eq:eq_Delta1}) with respect to $\Delta$ and
taking into account that in this case $\delta\tau\ll1$, we derive
the formula
\begin{equation}
     v=\frac{\theta_{1}\theta_{2}d}{\theta_{1}(d-c)+
     \theta_{2}c}\,,
     \label{eq:velocity2}
\end{equation}
which shows that for $1\ll K\ll\max{K}$ the chain velocity does
not depend on the chain characteristics. Since $1<A/R< c/(d-c)$,
for a given ratio $A/R=\sigma$ ($\sigma>1$) the parameter $c/d$
must belong to the interval $(\sigma/(\sigma+1),1)$. On this
interval, $v$ is a decreasing function of $c/d$, and according to
Eq.~(\ref{eq:velocity2}) $v\to\theta_{1}2\sigma/(\sigma+1)$ as
$c/d\to\sigma/(\sigma+1)$, and $v\to\theta_{1}$ as $c/d\to1$.

\section{Conclusions}

We have studied the ground state, vibration properties and
deterministic transport of a chain of charged particles
interacting via both the Coulomb interaction and a repulsive
interaction, which is characterized by the power $r$. We have
shown that a chain with equidistant particles has  minimum energy
for $r>r_{2}\approx2.799$. For a such chain, we have calculated
its energy and period as functions of $r$, derived and analyzed in
detail its phonon spectrum, and established that it is stable with
respect to small displacements of particles from their equilibrium
positions only if $r>r_{1}\approx 2.779$.

We have studied also the deterministic transport of a chain in the
case when the particles are acted upon by both a longitudinal
alternating electric field and a force field generated by a
ratchet potential. We have derived the set of equations that
describes the chain transport, analyzed the general properties of
its solution, and calculated the chain velocity in the overdamped
limit. If the period of the electric field is large enough, then
the chain velocity strongly depends on the interparticle
interactions. Otherwise, i.e., if the period of the electric field
is small enough, the chain velocity does not depend on them.

\section*{ACKNOWLEDGMENT}

We are grateful to Werner Horsthemke for a critical reading of the
manuscript and his valuable comments.

\appendix
\section{Another form for \mbox{\boldmath $U_0(\alpha,
\beta)$}}

Using the representation\cite{BaEr53}
\begin{equation}
        \psi(z)=\lim_{k\to\infty}\left[\ln k - \frac{1}{z} -
        \sum_{n=1}^{k}\frac{1}{n+z}\right]
        \label{eq:def_psi}
\end{equation}
for $\psi(z)=d\ln\Gamma(z)/dz$ ($\Gamma(z)$ is the gamma function)
and the definition
\begin{equation}
        \zeta(r,z)=\sum_{n=0}^{\infty}\frac{1}{(n+z)^{r}}
        \label{eq:def_zeta}
\end{equation}
of the generalized zeta function $\zeta(r,z)$, we find
\begin{equation}
        \sum_{n=1}^{\infty}\left[\frac{1}{n-\alpha}+\frac{1}
        {n+\alpha}-\frac{2}{n}\right]=2\psi(1)-\psi(\alpha)
        -\psi(-\alpha)
        \label{eq:sumA1}
\end{equation}
and
\begin{equation}
        \sum_{n}\frac{1}{|n+\alpha|^{r}}
        =\zeta(r,\alpha)+\zeta(r,1-\alpha).
        \label{eq:sumA2}
\end{equation}
Substituting Eqs.~(\ref{eq:sumA1}) and (\ref{eq:sumA2}) into
Eq.~(\ref{eq:energy_equil1}) and taking into account that
$\psi(-\alpha)=\psi(1-\alpha)+1/\alpha$,\cite{BaEr53} we obtain
the desired representation
\begin{eqnarray}
        U_{0}(\alpha,\beta)&=&-\frac{1}{2\beta}[2\psi(1)-
        \psi(\alpha)-\psi(1-\alpha)]
        \nonumber\\
        &&+\frac{1}{2\beta^{r}}[2\zeta(r)+\zeta(r,\alpha)
        +\zeta(r,1-\alpha)].
        \nonumber\\
        \label{eq:energy_0A1}
\end{eqnarray}

\section{Integral representation for \mbox{\boldmath $\Phi_3
(\kappa)$} and \mbox{\boldmath $\Psi_3(\kappa)$}}

We proceed from the standard series\cite{PrBrMa89}
\begin{equation}
     \sum_{p=1}^{\infty}\frac{\cos(pz)}{p}=-\ln\left(2\sin
     \frac{z}{2}\right)
     \label{eq:standard}
\end{equation}
$(0\leq z \leq\pi)$. Integrating Eq.~(\ref{eq:standard}) twice, we
find
\begin{equation}
     \sum_{p=1}^{\infty}\frac{1-\cos(pz)}{p^{3}}=-\int_{0}^{z}dx
     \int_{0}^{x}dy\ln\left(2\sin\frac{y}{2}\right),
     \label{eq:doubl}
\end{equation}
and using the definition (\ref{eq:defPhi}), we obtain
\begin{equation}
     \Phi_{3}(\kappa)=\zeta(3)+\int_{0}^{2\pi\kappa}
     dx\int_{0}^{x}dy\ln\left(2\sin\frac{y}{2}\right).
     \label{eq:intPhi}
\end{equation}

To find the integral representation for $\Psi_{3}(\kappa)$, we use
the identity
\begin{equation}
     \sum_{p=1}^{\infty}\frac{\cos(pz)}{p}=
     \sum_{p=1}^{\infty}\frac{\cos(2pz)}{2p}+
     \sum_{p=1}^{\infty}\frac{\cos[(2p-1)z]}{2p-1}
     \label{eq:identity}
\end{equation}
and the series (\ref{eq:standard}) to obtain
\begin{equation}
     \sum_{p=1}^{\infty}\frac{\cos[(2p-1)z]}{2p-1}=-\frac{1}{2}\ln
     \left(\tan\frac{z}{2}\right).
     \label{eq:sum5}
\end{equation}
Integrating Eq.~(\ref{eq:sum5}) by the same way as
Eq.~(\ref{eq:standard}) and using the definition
(\ref{eq:defPsi}), we find
\begin{equation}
     \Psi_{3}(\kappa)=\frac{7}{8}\zeta(3)+\frac{1}{2}\int_{0}^
     {\pi\kappa}dx\int_{0}^{x}dy\ln\left(\tan\frac{y}{2}\right).
     \label{eq:intPsi}
\end{equation}

\section{Derivation of Equation~(\ref{eq:eq_Delta})}

Let $t_{k}$ ($k=0,1,...,2K$, $t_{0}=0$, $t_{2K}\leq\tau$) and
$\bar{t}_{k}$ ($k=0,1,...,2K-1$, $t_{k} < \bar{t}_{k} < t_{k+1}$)
be the times at which the positive charges reach the minima and
maxima of $V(x)$ during the half-period $\tau$, i.e.,
$w(t_{k})=-\Delta/2+kd$ and $w(\bar{t}_{k})=-\Delta/2+kd+c$.
Solving Eq.~(\ref{eq:eq_basic}), we find
\begin{equation}
     w(t)=\frac{\theta_{1}}{\delta}-\left(\frac{\theta_{1}}
     {\delta}+\frac{\Delta}{2}-kd\right)e^{-\delta(t-t_{k})}
     \label{eq:sol1}
\end{equation}
for $t\in(t_{k},\bar{t}_{k})$, and
\begin{equation}
     w(t)=\frac{\theta_{2}}{\delta}-\left(\frac{\theta_{2}}
     {\delta}+\frac{\Delta}{2}-kd-c\right)e^{-\delta(t-
     \bar{t}_{k})}
     \label{eq:sol2}
\end{equation}
for $t\in(\bar{t}_{k},t_{k+1})$. At $t=\bar{t}_{k}$,
Eq.~(\ref{eq:sol1}) leads to
\begin{equation}
     e^{\delta(\bar{t}_{k}-t_{k})}=\frac{\theta_{1}
     /\delta+\Delta/2-kd}{\theta_{1}/\delta+\Delta/2-kd-c},
     \label{eq:rel1}
\end{equation}
and Eq.~(\ref{eq:sol2}) at $t=t_{k+1}$ yields
\begin{equation}
     e^{\delta(t_{k+1}-\bar{t}_{k})}=\frac{\theta_{2}/
     \delta+\Delta/2-kd-c}{\theta_{2}/\delta+\Delta/2-kd-d}.
     \label{eq:rel2}
\end{equation}
Multiplying Eqs.~(\ref{eq:rel1}) and (\ref{eq:rel2}), substituting
the result
\begin{equation}
     e^{\delta(t_{k+1}-t_{k})}=\frac{\theta_{1}
     /\delta+\Delta/2-kd}{\theta_{1}/\delta+\Delta/2-kd-c}\,
     \frac{\theta_{2}/\delta+\Delta/2-kd-c}
     {\theta_{2}/\delta+\Delta/2-kd-d}
     \label{eq:rel3}
\end{equation}
into the identity $e^{\delta t_{2K}}=\prod_{k=0}^ {2K-1}
e^{\delta(t_{k+1}-t_{k})}$, and using the formula
$\prod_{k=1}^{2K}(x-k)= \Gamma(x)/\Gamma(x-2K)$ and condition
$\Delta= 2Kd$, we obtain Eq.~(\ref{eq:eq_Delta}).

\cleardoublepage

\begin{figure}[h]
     \centering
     \includegraphics{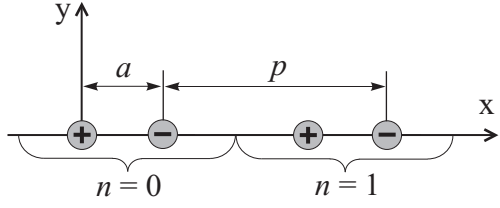}
     \caption{\label{fig1}Part of the equilibrium chain of charged
     particles.}
\end{figure}

\

\begin{figure}[h]
     \centering
     \includegraphics{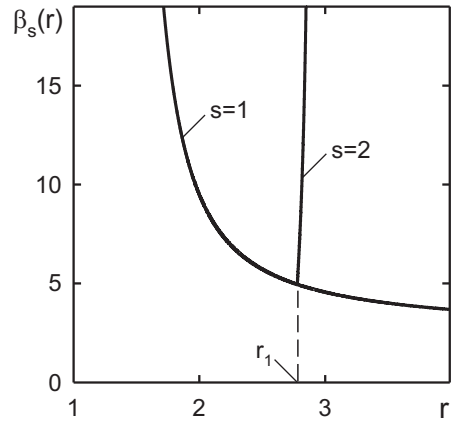}
     \caption{\label{fig2}Plots of $\beta_{1}(r)$ and
     $\beta_{2}(r)$; for the third solution $\beta_{3}(r)=\infty$.
     The function $\beta_{2}(r)$ increases from $\beta_{2}(r_{1})
     =\beta_{1}(r_{1}) \approx 4.959$ to $\beta_{2}(3)=\infty$.}
\end{figure}

\

\begin{figure}
     \centering
     \includegraphics{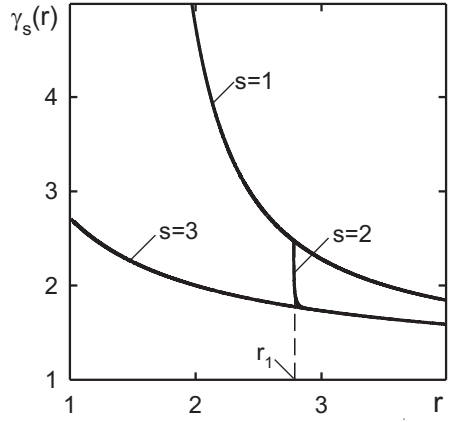}
     \caption{\label{fig3}Plots of $\gamma_{s}(r)$. All functions
     $\gamma_{s}(r)$ decrease monotonically, and $\gamma_{1}
     (r_{1})=\gamma_{2}(r_{1}) \approx 2.420$, $\gamma_{2}
     (3)=\gamma_{3}(3)=\sqrt{3}$.}
\end{figure}

\

\begin{figure}
     \centering
     \includegraphics{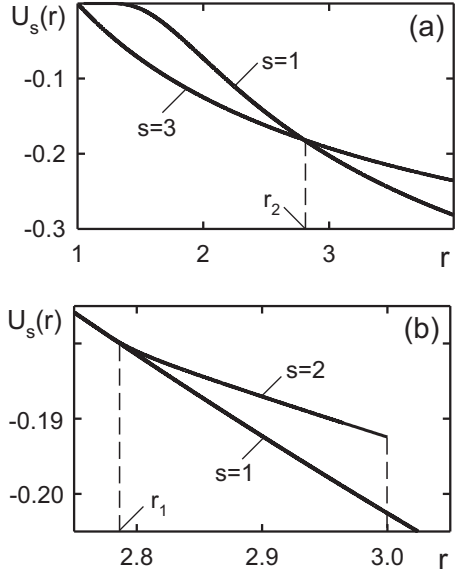}
     \caption{\label{fig4}Plots of $U_{1}(r)$ and $U_{3}(r)$
     [figure (a)], and $U_{1}(r)$ and $U_{2}(r)$ [figure (b)]. The
     functions $U_{3}(r)$ and $U_{2}(r)$ are very close to each
     other [$U_{3}(r)< U_{2}(r)$ for $r_{1}\leq r<3$, $U_{3}(3)
     =U_{2}(3)=-1/3\sqrt{3}$, and $U_{2}(r_{1})-U_{3}(r_{1})
     \approx 0.001$], therefore the plot of $U_{3}(r)$ in figure
     (b) is not shown.}
\end{figure}

\

\begin{figure}
     \centering
     \includegraphics{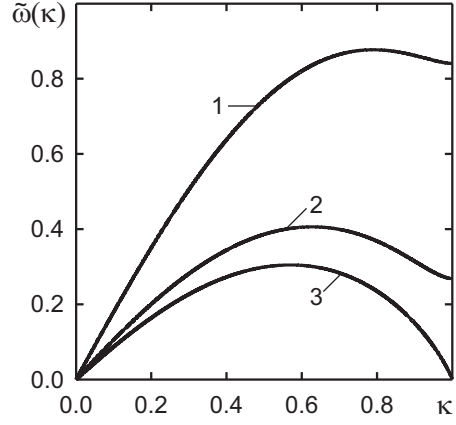}
     \caption{\label{fig5}Dependence of the phonon spectrum
     $\tilde{\omega}(\kappa)$ on $\kappa$ for $r=4$ (curve 1),
     $r=3$ (curve 2), and $r=r_{1}$ (curve 3). Since for $r=r_{1}$,
     the function $|\tilde{\omega}'(\kappa)|$ goes
     to infinity very slowly [as $\sqrt{|\ln(1-\kappa)}|$]
     as $\kappa\to1$, this feature of the spectrum is not
     represented on the chosen scale of the $\kappa$ axis.}
\end{figure}

\

\begin{figure}
     \centering
     \includegraphics{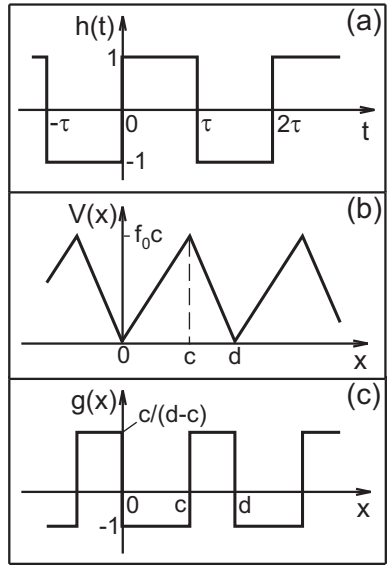}
     \caption{\label{fig6}Illustrative plots of the reduced
     alternating electric field $h(t)$ [figure (a)], the ratchet
     potential $V(x)$ [figure (b)], and the reduced force field
     $g(x)$ [figure (c)].}
\end{figure}
\end{document}